\documentclass[12pt]{article}
\usepackage{graphicx}
\usepackage{subfigure}
\newcommand{\beq}{\begin{equation}}
\newcommand{\eeq}{\end{equation}}
\newcommand{\bea}{\begin{eqnarray}}
\newcommand{\eea}{\end{eqnarray}}

\newcommand{\ra}{\right\rangle}
\newcommand{\la}{\left\langle}

\def\sq{{\vbox {\hrule height 0.6pt\hbox{\vrule width 0.6pt\hskip 3pt
   \vbox{\vskip 6pt}\hskip 3pt \vrule width 0.6pt}\hrule height 0.6pt}}}

\begin{document}
\begin{titlepage}
\begin{flushleft}
       \hfill                      {\tt hep-th/0612xxx}\\
       \hfill                       FIT HE - 06-03 \\
       \hfill                       KYUSHU-HET ** \\
       \hfill                       Kagoshima HE - 06-3 \\
\end{flushleft}
\vspace*{3mm}
\begin{center}
{\bf\LARGE Flavor quarks in AdS${}_4$ 
and gauge/gravity correspondence}

\vspace*{5mm}
\vspace*{12mm}
{\large Kazuo Ghoroku\footnote[2]{\tt gouroku@dontaku.fit.ac.jp},
Masafumi Ishihara\footnote[3]{\tt masafumi@higgs.phys.kyushu-u.ac.jp},
Akihiro Nakamura\footnote[4]{\tt nakamura@sci.kagoshima-u.ac.jp}
}\\
\vspace*{2mm}

\vspace*{2mm}

\vspace*{4mm}
{\large ${}^{\dagger}$Fukuoka Institute of Technology, Wajiro, 
Higashi-ku}\\
{\large Fukuoka 811-0295, Japan\\}
\vspace*{4mm}
{\large ${}^{\ddagger}$Department of Physics, Kyushu University, Hakozaki,
Higashi-ku}\\
{\large Fukuoka 812-8581, Japan\\
\vspace*{4mm}
{\large ${}^{\S}$Department of Physics, Kagoshima University, Korimoto 1-21-35,Kagoshima 890-0065, Japan\\}}

\vspace*{10mm}
\end{center}

\begin{abstract}
The non-perturbative properties of the gauge theories in the AdS${}_4$
are studied in the dual supergravity by including light flavor quarks, 
which are introduced by a D7 brane embedding.
Contrary to the cases of Minkowski
and dS${}_4$, the dilaton does not play any important dynamical role
in the AdS${}_4$ case, and the characteristic properties like 
the quark confinement and the chiral symmetry breaking are realized
mainly due to the geometry AdS${}_4$.
The possible hadron spectra 
are also examined, and we find that 
the meson spectra are well described by the formula given
by the field theory in AdS${}_4$, but the characteristic mass scale
is modified by the gauge interactions for exited states.

\end{abstract}
\end{titlepage}

\section{Introduction}

Recently, based on the gauge/gravity correspondence \cite{MGW},
many non-perturbative properties of Yang-Mills
theories with quarks have been uncovered in terms of
the  superstring theory \cite{KK}-\cite{GSUY}.   
In these approaches, the flavor quarks are introduced by embedding 
probe D7 brane(s) in an appropriate bulk 10d background, 
which could describe the QCD type of Yang-Mills theory,
and many successful results have been obtained for the properties 
of quarks and their bound states.  
 

Up to now, almost all 
efforts have been devoted to the study of the gauge theory
in the 4d Minkowski space-time. Previously, we extended our analysis to the
gauge theory in 4d de Sitter space, dS${}_4$, and many interesting properties
are found \cite{GIN,H}.
Here we furthermore extend the analysis 
into the case of 4d anti-de Sitter space-time, 
AdS${}_4$, which is embedded in the 10d bulk. The quantum 
field theory in the AdS${}_4$ has been studied in \cite{AIS}, and the discrete
mass spectra of scalar field has been found under the reflective boundary
conditions for the universal covering space CAdS. The same mass formula 
for the massive spin two states, which corresponds to the glueball states
in the gauge theory, are found from 5d AdS analysis by 
demanding the normalizability of the wave-function
in the fifth direction \cite{KR}. This implies the usefulness of the 
holographic approach to the gauge theory in the AdS${}_4$. In \cite{KR}, 
the existence of the massive graviton in AdS${}_4$ has been 
pointed out in a
three brane where gravity couples with CFT. And this observation has been
further studied by Porrati as a ``Higgs phenomena'' driven by dynamics in
the geometry AdS${}_4$ \cite{Poratti,BL}.

Here we concentrate our attention on
the properties of flavor quarks which couple to the CFT embedded 
in AdS${}_4$, which would give a strong constraint on the dynamics of
quarks interacting with deformed CFT as seen in the case of the gravity. 
The analysis is performed according to the
formulation given before for the case of dS${}_4$ \cite{GIN}.


\vspace{.2cm}
In section 2, we give the setting of our model for the present study. 
In section 3, the potential between quark and anti-quark
and the effective quark mass are studied through the Wilson 
Polyakov loop estimate. 
In section 4, the embedding of the D7-brane and the chiral symmetry breaking
are discussed. In section 5,
the possible bound state for the meson and baryon are
discussed. The meson spectra are compared with the formula given by the
field theory in AdS${}_4$. The summary is given in the final section.

\section{Background geometry}
We start from the type IIB supergravity with the following bosonic action,
\beq
 S={1\over 2\kappa^2}\int d^{10}x\sqrt{-g}\left(R-
{1\over 2}(\partial \Phi)^2+{1\over 2}e^{2\Phi}(\partial \chi)^2
-{1\over 4\cdot 5!}F_{(5)}^2
\right), \label{10d-action}
\eeq
where other fields are neglected since we need not them, and 
$\chi$ is Wick rotated. By taking the
ansatz for $F_{(5)}$, 
$F_{m_1\cdots m_5}=-\sqrt{\Lambda}/2~\epsilon_{m_1\cdots m_5}$ 
and $F_{\alpha_1\cdots \alpha_5}=\sqrt{\Lambda}/2~\epsilon_{\alpha_1\cdots \alpha_5}$
\cite{KS2,LT,GGP}, and for the 10d metric as $M_5\times S^5$ or
$ds^2=g_{mn}dx^mdx^n+g_{\alpha\beta}dx^{\alpha}dx^{\beta}$ \footnote{
Here $m,n$ denote $\mu,\nu=0\sim 3$ and the direction of $r$, 
the fifth coordinate. And $\alpha, \beta=5\sim9$.
}, the equations of motion are given as
\bea
 && R_{mn}=-\Lambda g_{mn} 
+{1\over 2}\partial_m \Phi\partial_n \Phi
   -{1\over 2}e^{2\Phi}\partial_m \chi\partial_n \chi \, , \label{5metric} \\
  &{}&{1\over \sqrt{-g}}\partial_m\left(\sqrt{-g}g^{mn}\partial_n\Phi
\right)=-e^{2\Phi}\partial_m \chi\partial_n \chi g^{mn} \, , \label{phieq} \\
  &{}&{1\over \sqrt{-g}}\partial_m\left(\sqrt{-g}g^{mn}e^{2\Phi}\partial_n\chi
\right)=0 \, , \label{chi-eq} \\
  && ~~~~~~~~~~~~~~~~~~R_{\alpha\beta}=\Lambda g_{\alpha\beta} \, .
\eea
Using the ansatz, 
\beq 
\chi=-e^{-\Phi}+\chi_0  \label{ansatz}
\eeq
where $\chi_0$ is a constant, the equation (\ref{5metric})
is written as 
\beq
R_{mn}=-\Lambda g_{mn}. \label{metric-eq-0}
\eeq
We notice that this equation (\ref{metric-eq-0}) is independent of the fields
$\Phi$ and $\chi$. This implies the usual 5d AdS solution, and we know that
the four dimensional slice perpendicular to the fifth coordinate $r$ is taken
as AdS${}_4$ as well as Minkowski or dS${}_4$. Here we consider the case of
AdS${}_4$ embedded in AdS${}_5$.

\vspace{.3cm}
After adopting the solution AdS${}_4$ embedded in AdS${}_5$ for the metric,
$\Phi$ and $\chi$ are obtained from (\ref{phieq}) and (\ref{ansatz}).
Thus we obtain the following solution,
$$ 
ds^2_{10}=G_{MN}dX^{M}dX^{N} ~~~~~~~~~~~~~~\hspace{6.5cm}
$$ 
\beq
=e^{\Phi/2}
\left\{
{r^2 \over R^2}A^2\left(-dt^2+a(t)^2\gamma(x)^2 (dx^i)^2\right)+
\frac{R^2}{r^2} dr^2+R^2 d\Omega_5^2 \right\} \ , 
\label{finite-c-sol}
\eeq 
\beq
e^\Phi= 1+\frac{q}{6}\left(1-{1-(r_0/r)^2\over (1+(r_0/r)^2)}\left[
1+{2(r_0/r)^2\over (1+(r_0/r)^2)^2}\right]\right) \ , 
\quad \chi=-e^{-\Phi}+\chi_0 \ ,
\label{dilaton}
\eeq
\beq
  A=1+({r_0\over r})^2, \quad a(t)={R^2\over 2r_0}\sin(2{r_0\over R^2} t), 
\quad \gamma(x)={1\over 1-x_ix^i/(4\tilde{r}_0^2)}\ , \label{scale}
\eeq
where $M,~N=0\sim 9$, $R=\sqrt{\Lambda}/2=(4 \pi N_c)^{1/4}$
and $\tilde{r}_0$ is an arbitrary scale factor. We comment on other two 
integration constants, $r_0$ and $q$. 
First, $r_0$ has nothing to do with the horizon since there is no horizon
in the present solution and $r$ is considered in the all
region of $0<r<\infty$ contrary to the dS${}_4$ case.
This parameter $r_0$ is however important since
it is related to the 4d cosmological constant $\lambda$, which is negative, as follows
\beq
  -\lambda=4{r_0^2\over R^4}.
\eeq
And $q$ is a constant which corresponds to the gauge fields
condensate~\cite{GY} defined as the vacuum expectation value (VEV) 
of gauge field squared as, 
\beq
  q\equiv \tilde{q}/r_0^4=\la F_{\mu\nu}^2\ra/r_0^4.  \label{gc}
\eeq
We assume that $\la F_{\mu\nu}^2\ra=\tilde{q}$ is finite, 
and other field configurations are set to be zero.

\vspace{.3cm}
{Our model is based on type IIB supergravity, and we solved the equations
of motion with dilaton and axion with the ansatz (\ref{ansatz}). In this case,
the condition for the supersymmetry of the solution is reduced to the
presence of the Killing spinor $u$ \cite{KS2,GGP}
which satisfies the following equation
in terms of the 10d gamma matrices $\Gamma_{M}$,
\beq
 \delta\Psi_{M}=\left(D_{M}-{\sqrt{\Lambda}\over 4}\Gamma_{M}\right)u=0\, ,
\label{gravitino}
\eeq
where $\Psi_{M}$ denotes the gravitino and 
$D_{M}=\partial_M+{1\over 4}\omega^{AB}_M\Gamma_A\Gamma_B$. 
The equation for the dilatino, 
$\delta\lambda=0$,
is satisfied due to the ansatz (\ref{ansatz}). The $S^5$ part 
of (\ref{gravitino}) is obtained
as in the case of AdS${}_5\times S^5$ \cite{LPR}.
For the AdS${}_5$ part the condition is modified due to the metric of
AdS${}_4$. In this case, the covariant derivatives are written as,
\bea
 D_ru&=&\partial_ru \\
 D_0u&=&\left(\partial_0-{\mu\over 2} \left({1-(r_0/r)^2\over 1+(r_0/r)^2}
           \right)
                   \Gamma_0\Gamma_r\right)u \\
 D_iu&=&\left(\partial_i
      -{\gamma(x)\over 2}\partial_0a(t)\hat{\Gamma}_i\hat{\Gamma}_0
      +{1\over 2\gamma(x)}\partial_j\gamma(x)\hat{\Gamma}_i\hat{\Gamma}_j 
        \right.                \nonumber \\
          ~~~~~&&\left. -{\mu\over 2a(t)\gamma(x)}
              \left({1-(r_0/r)^2\over 1+(r_0/r)^2}\right)
                    \Gamma_i\Gamma_r\right)u\, , \label{covariant-deri} 
\eea
where $\hat{\Gamma}_M$ denotes the local Lorentz constant gamma matrix. From
these, we find easily, for finite $r_0$,
that there is no Killing spinor, the solution of (\ref{gravitino}), 
of the form 
$u=f(y,x^{\mu})u_0$, where $u_0$ is some constant spinor. In other words,
there is no supersymmetry in the present case.
We notice that
the supersymmetry is also broken by our D7 brane embedding 
since the $\kappa$ symmetry of the D7 brane action is lost \cite{GY} when
the chiral symmetry is broken.}

\vspace{.3cm}
In this model, the four dimensional slice represents the AdS${}_4$
universe characterized by the 4d cosmological constant $\lambda$. Then the 
bulk gravity describes the gauge theory in the AdS${}_4$.
The gauge coupling constant in the present model is finite in the both
limits of ultraviolet and infrared,
\beq
 g_{\rm YM}^2=e^{\Phi}=\left\{\begin{array}{ll}
      1+{q\over 3}-q\left({r\over r_0}\right)^4  & \mbox{for}\ r\to 0 
           \ \mbox{(IR limit)}\\
      1+q\left({r_0\over r}\right)^4 & \mbox{for}\ r\to \infty \ \mbox{(UV limit)}
                   \end{array}\right. .
\eeq
Meanwhile, in the limit of Minkowski space or the limit 
of $r_0\to 0$, we have
$ g_{\rm YM}^2=1+{\tilde{q}\over r^4}$,
which is equivalent to the supersymmetric solution given previously \cite{GY}.
In this case, quark is confined since the YM coupling constant
becomes very strong in the infrared limit, $r\to 0$ for $\tilde{q}>0$. 
But, for finite $r_0$ or in the AdS${}_4$ space-time, in the limit 
$r\to 0$, we find 
$ g_{\rm YM}^2\to 1+{q\over 3}$, and it is not enough large 
for the quark confinement.
However, the confinement is seen in this case due to the behavior of the
warp factor
$A(r)$ since it diverges at $r=0$ for finite $r_0$. This is
assured by the calculation
of the Wilson loop. Then the result for the quark confinement 
is independent of the value of $q$,
which is needed in the Minkowski space-time for the quark confinement, 
in the AdS${}_4$.

\section{ Wilson Loop and Quark Confinement} 
We study here the quark confinement through the Wilson-Polyakov loop
in $SU(N)$ gauge theory defined as, 
$ W={1\over N} \textrm{Tr} P e^{i\int A_0 dt}$ .
Then the quark-antiquark potential $V_{q\bar{q}}$ is derived from
the expectation value of a parallel Wilson-Polyakov loop. 
 From the dual gravity side, it is represented as
\beq
    \langle W\rangle  \sim e^{-S} , \label{wstr}
\eeq
in terms of the Nambu-Goto action 
\beq
   S=- \frac{1}{2 \pi \alpha'} 
\int d\tau d\sigma \sqrt{-\textrm{det}\, h_{ab}} , 
\eeq
with the induced metric
$
    h_{ab}=G_{\mu\nu}\partial_a X^{\mu}\partial_b X^{\nu} 
$. 
The string world-sheet is parameterized by $\sigma$, $\tau$.

We examine quark-antiquark potentials
in the background given above. 
To study possible static string configurations 
of a pair of quark  and anti-quark,
we choose as $X^0=t=\tau$ and $X^1=x^1=\sigma$, then 
the Nambu-Goto Lagrangian in the background (\ref{finite-c-sol}) 
becomes
\beq
   L_{\textrm{\scriptsize NG}}=-{1 \over 2 \pi \alpha'}\int d\sigma ~e^{\Phi/2}
   A(r)\sqrt{r'{}^2
        +\left({r\over R}\right)^4 \left(A(r) a(t) \gamma(x)\right)^2 
} ,
 \label{ng}
\eeq
where the prime denotes the derivative with respect to $\sigma$.
The test string has two possible configurations:
(i) a pair of parallel string, which connects the D7 and D3 branes,
and (ii) a U-shaped string whose two end-points are on the D7 brane.

\vspace{.3cm}
First, we study the 
configuration (i).  In this case, the parallel
strings have no correlation each other, then
the total energy of this configuration 
is two times of one effective quark mass,
$\tilde{m}_q$. It is given by a string configuration which stretches 
between  $r=0$ and $r_{\rm max}$, which denote the position of D3 and 
D7 branes respectively. So we can take as $r=\sigma$, then we obtain
\beq
   \tilde{m}_q=
   {1\over 2\pi \alpha'}\int_{0}^{r_{\textrm{\scriptsize max}}}
     dr ~e^{\Phi/2}A(r)  \ . \label{dynamicalmass}
\eeq
Near $r=0$, the integrand is approximated as 
$e^{\Phi/2}A(r)\sim \sqrt{1+{q\over 3}}\left({r_0\over r}\right)^2$,
then we find that $\tilde{m}_q$ diverges as
\beq
  \tilde{m}_q\sim \sqrt{1+{q\over 3}}~\left.{r_0^2\over r}\right|_{r\to 0}
               \to\infty.
\eeq
Thus the configuration of (i) can not be
realized. In other words, the quark should confined.  
And this result is satisfied even if $q=0$, then the confinement
is caused only by the gravitational effect. 

\vspace{.3cm}
The next check of the quark confinement is to see the area law through
the U-shaped configuration. The energy
of this configuration is obtained from (\ref{ng}) as
\beq
 E=-L_{\rm NG}={1 \over 2 \pi \alpha'}\int d\tilde{\sigma}~e^{\Phi/2}
   A(r)\sqrt{(\partial_{\tilde{\sigma}}r)^2
        +\left({r\over R}\right)^4 \left(A(r) \right)^2 } \, 
\eeq
\beq
={1\over 2\pi \alpha'} \int d\tilde{\sigma}~ n~
        \sqrt{1+ \left({R^2\over r^2 A}\partial_{\tilde{\sigma}}r
               \right)^2}\ , \label{W-energy}
\eeq
where 
\beq
 n=e^{\Phi/2}\left({r A\over R}\right)^2 \, ,
\eeq
and 
\beq
  \tilde{\sigma}=a(t)\int d\sigma\gamma(\sigma)
       =a(t)\int d\sigma{1\over 1-\sigma^2/4} \, .
\eeq
Here the coordinates of the string action are set as $x^1=\sigma, x^2=x^3=0$.
In this case, we obtain
\beq
  \tilde{\sigma}=4a(t)\tanh^{-1}\left(\tan \left({\theta\over 2}\right)\right)
  \, , \quad \sigma=2\cos(\theta)\, ,
\eeq
where $0\leq\theta <\pi/2$, and we find $0\leq\tilde{\sigma} <\infty$.
The physical distance between the quark and 
anti-quark is measured by $\tilde{\sigma}$ rather than $\sigma$ which is
restricted as $-2<\sigma<2$ in our definition of the metric given by 
(\ref{finite-c-sol}).

Then the distance between the quark and the anti-quark at $t=0$ is given
as
\beq
 \tilde{L}=2\int_{\tilde{\sigma}_{\rm min}}^{\tilde{\sigma}_{\rm max}}
   d\tilde{\sigma}=2({\tilde{\sigma}_{\rm min}}-{\tilde{\sigma}_{\rm max}})\ ,
\eeq
where
$
    \tilde{\sigma}_{\rm min}=\tilde{\sigma}(r_{\rm min})\ ,~
 \tilde{\sigma}_{\rm max}=\tilde{\sigma}(r_{\rm max})\ ,
$ 
and $r_{\rm min}$ is determined as 
$\partial_{\tilde{\sigma}}r|_{r_{\rm min}}=0$. So $r_{\rm min}$ gives
a midpoint of the U-shaped string. On the other hand, $r_{\rm max}$ is
given by the position of the D7 brane, the end points of the U-shaped 
string.

\vspace{.3cm}
Before giving the numerical estimation of the Wilson loop, we give an
analytic approximate estimation according to the method given by Gubser
\cite{Gubser}.
The Eq.~(\ref{W-energy}) can be approximately evaluated 
in terms of the classical solution for $r$, say
$r^{*}(\tilde{\sigma})$, which
minimizes $E$.
And it is approximated by the global minimum of the
function $n$ with respect to $r$. For this solution, we expect
$\partial_{\tilde{\sigma}}r^{*}\sim 0$ for a wide range of $\sigma$
for large $\tilde{L}$.
Then we obtain
\beq
 E\sim {n(r^{*})\over 2\pi \alpha'} \tilde{L}\ , \label{linear-P}
\eeq
and this implies the linear potential between quark and anti-quark.
For the case of small $q$, we obtain
\bea
 r^{*}&=&r_0\left(1+{q\over 16}+\cdots \right)\ ,\\ 
 n(r^{*})&=&|\lambda| ~R^2\left(1+{q\over 12}+\cdots \right)\ ,
\eea
where we notice $\lambda=-4 r_0^2/R^4$. After all we find the tension
$\tau_{U}$ of the U-shaped string configuration as
\beq
 \tau_{U}={|\lambda| ~R^2\over 2\pi \alpha'}\left(1+{q\over 12}+\cdots \right).
\label{tention}
\eeq
As expected, the tension is finite even if
$q=0$, and $\lambda$ is the essential factor to obtain the linear
potential or the quark confinement. The relation of (\ref{linear-P})
is explicitly shown by the numerical calculation.

 From the Lagrangian (\ref{ng}), we find the following relation,
\beq
     e^{\Phi/2}{1\over \sqrt{(r/R)^4 A^2(r)+(dr/d\tilde{\sigma})^2}}
    \left({r\over R}\right)^4 A^3(r)= h\ ,
\eeq
where $h$ denotes a constant of motion. Taking as 
$h=e^{\Phi/2}\left({r\over R}\right)^2 A^2(r)|_{r_{min}}$, we get
\bea
  &&  \tilde{L}=2R^2 \int_{r_{min}}^{r_{\textrm{\scriptsize max}}} dr~
      {1\over r^2 A(r)
        \sqrt{e^{\Phi(r)}r^4 A(r)^4 /
          \left(e^{\Phi(r_{min})}r_{min}^4A(r_{min})^4\right)-1}} , \label{len}
\\
  && E=
   {1\over \pi \alpha'} \int_{r_{min}}^{r_{\textrm{\scriptsize max}}}dr~
   {A(r)e^{\Phi(r)/2}\over 
     \sqrt{1-e^{\Phi(r_{min})}r_{min}^4 A(r_{min})^4/
             \left(e^{\Phi(r)}r^4 A(r)^4\right)}} . \label{energy}
\eea
\begin{figure}[htbp]
\vspace{.3cm}
\begin{center}
\includegraphics[width=10.0cm,height=5cm]{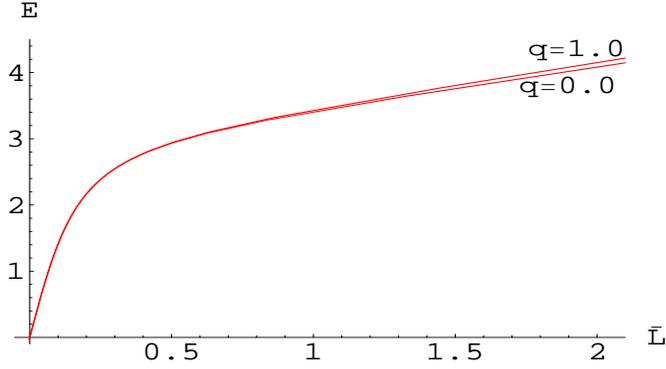}
\caption{Plots of $E$ vs $\tilde{L}$ at $q=1$ and $q=0$
for $\lambda=-4$, $R = 1\, (\mbox{GeV}^{-1})$, 
$r_{\textrm{\scriptsize max}}=10\, (\mbox{GeV}^{-1})$ 
and $\alpha'=1\, (\mbox{GeV}^{-2})$.
\label{el}}
\end{center}
\end{figure}
Figure~\ref{el} shows the dependence of the energy $E$
on the distance $\tilde{L}$ at finite cosmological constant $\lambda(=-4)$ 
for $q=1$ and $q=0$. In both cases, we find
the linear potential and the tension given by the Eq.~(\ref{tention})
is assured.
We should notice that this
confinement is realized even if $q=0$, so the gauge interactions
are not essential in this case.

\vspace{.3cm}
{ Here we notice that the parameter $r_{min}$ is introduced instead of $h$,
defined in (\ref{len}),
in the above calculation. When $r_{min}$ approaches to $r_0$, 
$f(r_{min})\equiv e^{\Phi(r_{min})}r_{min}^4 A(r_{min})^4$ approaches to
its minimum value. For small $q$, it is realized at 
$$r=r_0(1+q/16+\cdots)\equiv r_M.$$ Then, for $r_{min}=r_M$, $\tilde{L}$
diverges as
\beq
 \tilde{L}\sim -{2R^2\over r_M^2A(r_M)}\sqrt{{2f(r_M)\over f''(r_M)}}
        \log(r-r_M)\big|_{r\to r_M}
\eeq
Therefore, it is enough to consider $r_{min}$ in the range of 
$r_{min}>r_M$ in order to see the behavior of
the U-shaped Wilson loop configuration. On the other hand, we
can not find the U-shaped Wilson loop when we enter into the region of 
$r_{min}<r_M$. Instead, we find an upended shaped one in which $r_{min}$
is the top of the string. So the end points go to $r=0$ not to $r_{max}$.
We do not consider this unphysical configuration.
Therefore, the natural configuration is obtained when
the D7 brane position $r_{max}$ is taken
as $r_{max}>r_{min}>r_M$. This point is assured in the next section.}

\vspace{.3cm}
\section{D7 brane embedding and Chiral symmetry}
The flavor 
quarks are introduced by embedding D7 brane(s) in the following
rewritten background, 
\bea 
  ds_{10}^2&=& e^{\Phi/2}\left\{
{r^2 \over R^2}A^2\left(-dt^2+\gamma(x)^2a(t)^2(dx^i)^2\right) \right. \nonumber \\
 &+&\left.
\frac{R^2}{r^2}\left(d\rho^2+\rho^2d\Omega_3^2+(dX^8)^2+(dX^9)^2
\right) \right\} \ , 
\label{d7}
\eea
where $r^2=\rho^2+(X^8)^2+(X^9)^2$. Then
the induced metric for D7 brane is obtained as,
$$ 
ds^2_8=e^{\Phi/2}
\left\{
{r^2 \over R^2}A^2\left(-dt^2+\gamma(x)^2a(t)^2(dx^i)^2\right)+\right.
\hspace{3cm}
$$
\beq
\left.\frac{R^2}{r^2}\left((1+(\partial_{\rho}w)^2)d\rho^2+\rho^2d\Omega_3^2\right)
 \right\} \ , 
\label{D7-metric}
\eeq
where we set as $X^9=0$ and $X^8=w(\rho)$
without loss of generality due to the rotational invariance in
$X^8-X^9$ plane. 
Then, the effective D7 brane action is given as
\beq
S_{\rm D7}= -\tau_7 \int d^8\xi \left(e^{-\Phi}
    \sqrt{-\det\left({\cal G}_{ab}+2\pi\alpha' F_{ab}\right)}
      -{1\over 8!}\epsilon^{i_1\cdots i_8}A_{i_1\cdots i_8}\right)
\label{D7-action}
\eeq
where $F_{ab}=\partial_aA_b-\partial_bA_a$.
${\cal G}_{ab}= \partial_{\xi^a} X^M\partial_{\xi^b} X^N G_{MN}~(a,~b=0\sim 7)$
and $\tau_7=[(2\pi)^7g_s~\alpha'~^4]^{-1}$ represent the induced metric and
the tension of D7 brane respectively.
The eight form potential $A_{i_1\cdots i_8}$,
which is the Hodge dual to the axion, couples to the
D7 brane minimally. In terms of the Hodge dual field strength,
$F_{(9)}=dA_{(8)}$ \cite{GGP}, the potential $A_{(8)}$ is obtained. 

Taking the canonical gauge, we arrive at the following D7 brane
action,
\beq
S_{\rm D7} =-\tau_7~\int d^8\xi  \sqrt{\epsilon_3}\rho^3 \gamma(x)^3a(t)^3
\left(A^4
   e^{\Phi}\sqrt{ 1 + (w')^2 }-{\tilde{q}\over r^4}\right) 
\ ,
\label{D7-action-2}
\eeq
\vspace{.3cm}
Then the equation of motion for $w(\rho)$ is obtained as, 
\bea
   {w\over \rho+w~w'}
   \left[\Phi'-\sqrt{1+(w')^2}(\Phi+4\log A)'~\right]
\nonumber\\
   +{1\over \sqrt{1+(w')^2}}
\left[w'\left({3\over \rho}+(\Phi+4\log A)'\right)
+ {w''\over 1+(w')^2}\right]
   &=&0 ,
  \label{qeq}
\eea
where prime denotes the derivative with respect to $\rho$. 
By solving this
equation we find the embedded configuration of D7 brane. And from this
solution, the quark mass, $m_q$,
and the chiral condensate, $c=-\langle\bar{\Psi}\Psi\rangle$, are found through
its asymptotic form at large $\rho$ as,
\beq
   w(\rho) \sim m_q+{c\over \rho^2} ,  \label{asym}
\eeq
according to the gauge/gravity correspondence.
However we notice that the above asymptotic form, (\ref{asym}), 
must be modified. 
This is seen by expanding the Eq.~(\ref{qeq}) in terms of the power series
of $1/\rho^2$ by adding the power series \cite{GIN}, and we obtain
\beq
  w(\rho) \sim m_q+{c_0+4m_q r_0^2\log(\rho)\over \rho^2} ,  \label{asym2}
\eeq
This implies the chiral condensate receives quantum corrections in the
limit of $\rho\to 0$ as $c=c_0+4m_q r_0^2\log(\rho)$. In other words,
the conformal invariance of the 4d CFT is broken by
the added chiral multiplet for the flavor quarks.

\begin{figure}[htbp]
\vspace{.3cm}
\begin{center}
  \includegraphics[width=11.5cm]{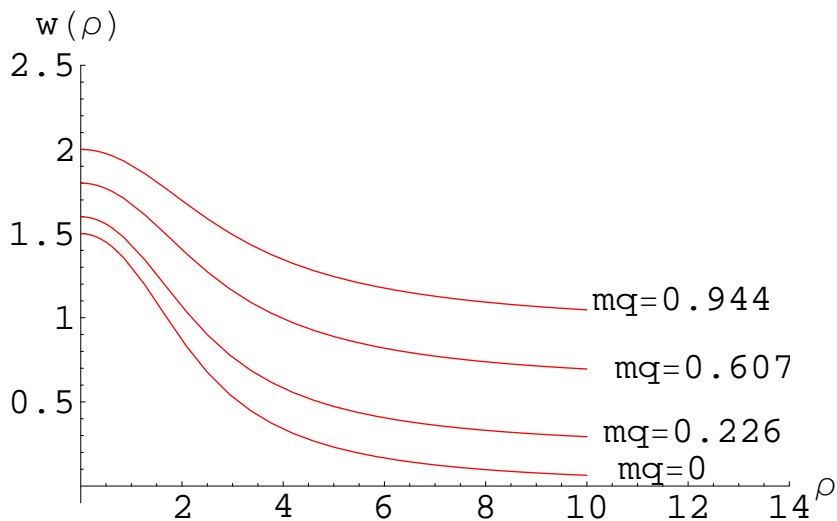}
\caption{Embedding solutions for $q=1.0$ and $r_0=1.0$. 
The solutions are drawn for several values of $m_q$, where we have taken 
as
$R=1$. \label{wq0fig}}
\end{center}
\end{figure}

\vspace{.3cm}
The solutions are obtained in terms of the above asymptotic form (\ref{asym2}),
and the results are shown in the Fig.~\ref{wq0fig} for several values of $m_q$
in the range $0\leq m_q<1$. We find $c>0$ for all the solutions. 
This is understood from the repulsive force between D3 and D7 branes.  
The potential of the D7 brane near the D3 branes is given from 
the D7 action as,
\beq
 V(w)=\tau_7\left(A^4~e^{\Phi}-{q\over r^4}-1\right) 
     =\tau_7\left\{\left(1+{q\over 3 r_0^4}\right)\left({r_0\over r}\right)^8
        +\cdots \right\}\, . \label{series-P}
\eeq
We find a strong repulsion near $r=0$.

\vspace{.3cm}
In the Fig.~\ref{wq0fig}, the solution of $m_q=0$ with $c>0$, which then breaks
the chiral symmetry, is
shown. On the other hand, the trivial solution with $w=0$ and $m_q=0$, which
preserves the chiral symmetry since
$c=0$, also exists. The behaviors of the two solutions are very different
near $\rho=0$ although they are very similar at large $\rho$.
We find for the trivial solution that the energy density of the D7 brane is
divergent like $1/\rho^5$ at $\rho=0$ since $V\sim 1/\rho^8$ near $\rho=0$. 
On the other hand, for the
solution of $m_q=0$ with $c>0$, the D7 energy density approaches to zero
like $\rho^3$ near $\rho=0$, and as a result 
we obtain a finite energy after $\rho$
integration in this case. Thus we can say that 
the chiral symmetry is spontaneously broken in the present case.

\vspace{.3cm}
{The second point to be addressed is that the solutions are far from the circle
$r\sim r_0$. This is consistent with the stability of the fluctuations of D7
brane since the U-shaped string configurations can not be stretched up 
to $r=r_0$ as mentioned in the previous section. }

\section{Possible hadron spectrum}
As shown in the previous section, the theory is in the quark
confined phase and the U-shaped string configuration exists. And the
tension parameter of the string configuration
depends on the cosmological constant, $\lambda$, and $q$. 
So the meson mass also depends on these parameters and ${m}_q$. However,
the $q$ dependence is rather small, and we expect that the mass spectra
of the mesons are largely constrained by the geometry AdS${}_4$.

We study these point by calculating the meson mass and
comparing it with the one given by the field theory in AdS${}_4$. 


\subsection{Meson spectrum}
The meson spectrum is obtained by solving the equations of motion
of the fields on the D7 brane. According to \cite{BGN, GIN}, firstly we
consider the fluctuations of the scalar mesons which are defined as,
$$X^9=\tilde\phi^9,\quad X^8=w(\rho)+\tilde\phi^8.$$
And writing the wave functions in the following factorized form,
$$\tilde\phi^k=\varphi^k(t,x^i)\phi_l^k(\rho){\cal Y}_l(S^3),\qquad(k=9,8)$$
where ${\cal Y}_l(S^3)$ denotes the spherical harmonic function on 
three dimensional sphere with the angular momentum $l$.   
Then we study the linearlized field equations for $\phi_l^9(\rho)$ 
and $\phi_l^8(\rho)$.

\vspace{.5cm}
\subsubsection{Solution for $w=0$ and $q=0$}

\begin{figure}[htbp]
\vspace{.3cm}
\begin{center}
\includegraphics[width=10.0cm,height=5cm]{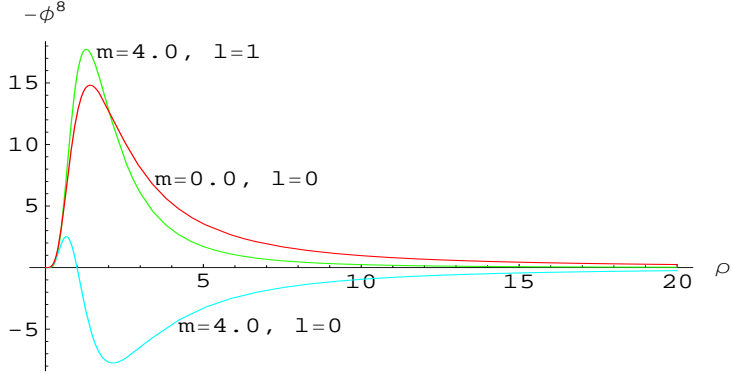}
\caption{The solutions for $w=q=0$. The one of (i) $m=0, l=0$, (ii) $m^2=-4\lambda, 
l=0$ and (iii) $m^2=-4\lambda, l=1$
are shown for $r_0=1.0$ and $R=1$, then $-\lambda=4$. The scale of the wave function for (iii) is expressed by multiplied by the extra factor 
$10^{-4}$ compared to the other two functions.
\label{susy-mass-fig}}
\end{center}
\end{figure}

At first, we consider the spectra of the mesons for a simple case where
$w=0$ and $q=0$. In this case, the equations of motion for $\phi_l^8$ and
$\phi_l^9$ have the following same form,
\beq
 \partial_{\rho}^2\phi_l+\left(
    {3\over \rho}-{8r_0^2\over A\rho^3}\right)\partial_{\rho}\phi_l
+\left[{m^2R^4\over A^2\rho^4}-{l(l+2)\over \rho^2}+{8r_0^2\over A\rho^4}
\right]\phi_l=0\, ,
  \label{phi}
\eeq
where $\phi_l$ denotes $\phi_l^8$ or $\phi_l^9$. The solution of this 
equation is obtained after some calculation as
\bea
 \phi_l&=&(r_0^2+\rho^2)^{-(3+\sqrt{9+\bar{m}^2})/2}\left\{c_1\rho^{4+l}
    F\left(\alpha,\alpha+l+1,l+2,-{\rho^2\over r_0^2}\right)\right. 
\nonumber \\
  &&\left.+c_2\rho^{2-l} 
    F\left(\alpha,\alpha-l-1,-l,-{\rho^2\over r_0^2}\right)\right\}\, ,
\label{super-sol}
\eea
\beq
  \alpha={1\over 2}(1-\sqrt{9+\bar{m}^2})\, , \quad \bar{m}=R^2m/r_0\, ,
\eeq
where $c_1$ and $c_2$ are arbitrary constants and
$F$ denotes the hypergeometric function. In order
to have a convergent solution $\phi$ at small $\rho$, we must take as $c_2=0$.
And from the condition at large $\rho$, $\alpha$ must be set as
\beq
 \alpha+l+1=-n\, , \quad n=0,1,2, \cdots
\eeq
The solution is expressed by the finite power series of $\rho$
due to this condition, and we find the normalizability of the wave-function.
Then we find the following result,
\beq
  m^2=-\lambda (l+n)(l+n+3)\, .  \label{mass-formula}
\eeq
This represents precisely the same spectrum with the one obtained by Avis, 
Isham and Storey \cite{AIS} in the AdS${}_4$ for the scalar fields. 
{This point is important since the mass scale $\lambda$ in the 
Eq.~(\ref{mass-formula}) is largely modified for the case of $w\neq 0$ and
$q\neq 0$ as shown in the next subsection. In the present case, $w=0$ and
$q=0$, then the quantum effect coming from the gauge interactions are all
suppressed. Then the above formula is obtained.}

\vspace{.3cm}
We show
the typical wave functions for $m^2=0$ and $m^2=-4\lambda$ in the 
Fig.~(\ref{susy-mass-fig}).
The mass eigenvalue $m=4$ is obtained by the two degenerate states,
$(n,l)=(0,1)$ and $(1,0)$, whose wave-functions are shown in the 
Fig.~(\ref{susy-mass-fig}). They have the different functional forms.
In the next sub-section, we can see the splitting of these degenerate mass
by introducing non-zero $w(\rho)$ and $q$. They therefore deviate from the
above mass formula (\ref{mass-formula}).

These solutions preserve the chiral symmetry since
the chiral condensate vanishes in the case of $w=0$. However, they are
not realized from the energy principle since the one of $w\neq 0$ with
the same boundary condition at $\rho=\infty$
is preferred, then the chiral symmetry is spontaneously broken. The mass
spectra in this realistic case $w\neq 0$ are examined by the numerical analyses
in the next sub-section, and we can see the details of the deviations of 
the meson spectra.

\vspace{.5cm}
\subsubsection{Solution for $w\neq 0$}
Next, we get the linearlized field equations for $\phi_l^9(\rho)$ 
and $\phi_l^8(\rho)$ for $w\neq 0$ as follows 
$$
 \partial_{\rho}^2\phi_l^9+
    {1\over L_0}\partial_{\rho}(L_0)\partial_{\rho}\phi_l^9
+(1+w'~^2)
\left[({R\over r})^4{m_9^2\over A^2}-{l(l+2)\over \rho^2}-2K_{(1)}
\right]\phi_l^9
$$
\beq
+(1+w'~^2)^{1/2}{1\over r}{{\partial\Phi}\over{\partial r}}\phi_l^9=0
  \label{phi9}
\eeq
\beq
 L_0=\rho^3e^{\Phi} A^4 {1\over\sqrt{1+w'~^2}},
  \quad    K_{(1)}={1\over e^{\Phi} A^4}\partial_{r^2}(e^{\Phi} A^4)
\eeq
and
$$
 \partial_{\rho}^2\phi_l^8+
    {1\over L_1}\partial_{\rho}(L_1)\partial_{\rho}\phi_l^8
+(1+w'~^2)\left[({R\over r})^4{m_8^2\over A^2}-{l(l+2)\over \rho^2}
     -2(1+w'~^2)(K_{(1)}+2w^2K_{(2)})
\right]\phi_l^8
$$
$$
+(1+w'~^2)^{3/2}\left[\left(2rK_{(1)}{{\partial\Phi}\over{\partial r}}
+{{\partial^2\Phi}\over{\partial r^2}}\right){{w^2}\over{r^2}}
+{{\partial\Phi}\over{\partial r}}{{\rho^2}\over{r^3}}\right]\phi_l^8
$$
\beq
 =-2{1\over L_1}\partial_{\rho}(L_0 w~w'K_{(1)})\phi_l^8  \label{phi8}
\eeq
\beq
 L_1={L_0\over {1+w'~^2}}, \quad
   K_{(2)}= {1\over e^{\Phi} A^4}\partial_{r^2}^2(e^{\Phi} A^4).
\eeq
Where four dimensional mass $m_9$ and $m_8$ are defined by 
\beq
-\sq_4\varphi^k=\ddot\varphi^k+3{{\dot a}\over a}\dot\varphi^k
-{1\over{a^2\gamma^3}}\partial_i\left(\gamma\partial_i\varphi^k\right)
=-m_k^2\varphi^k.\qquad(k=9,8)
   \label{mass}
\eeq
In deriving the above equations (\ref{phi9}) of $\phi_l^9$ and 
(\ref{phi8}) of $\phi_l^8$, we used 
\beq
 r^2=\rho^2+(\phi_l^8)^2+(\phi_l^9)^2+w^2+2w \phi_l^9
\eeq
But we should notice here that
the variable $r$ in the above field equations is understood as
$r^2=\rho^2+w^2$ since we are considering the linearlized equations.

\begin{figure}[htbp]
\vspace{.3cm}
\begin{center}
\subfigure[] {\includegraphics[angle=0,width=0.45\textwidth]{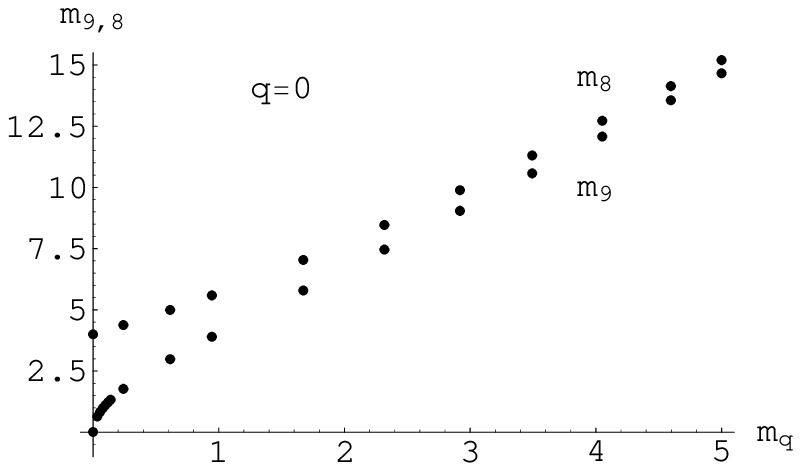}
            \label{mass-fig}}
\subfigure[] {\includegraphics[angle=0,width=0.45\textwidth]{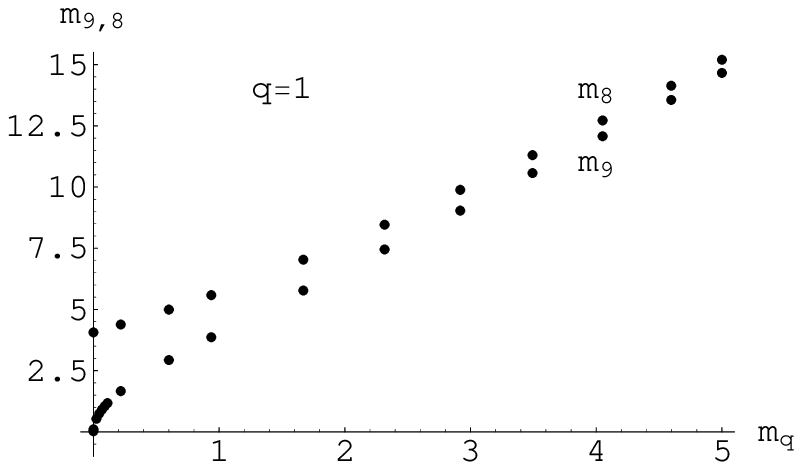}
            \label{mass2-fig}}
\caption{$m_{9,8}$ vs $m_q$ for $l=n=0$, $r_0=1.0$ and $R=1$ 
and (a) $q=0$, (b) $q=1$.
}
\end{center}
\end{figure}

\begin{figure}[htbp]
\vspace{.3cm}
\begin{center}
\subfigure[] {\includegraphics[angle=0,width=0.45\textwidth]{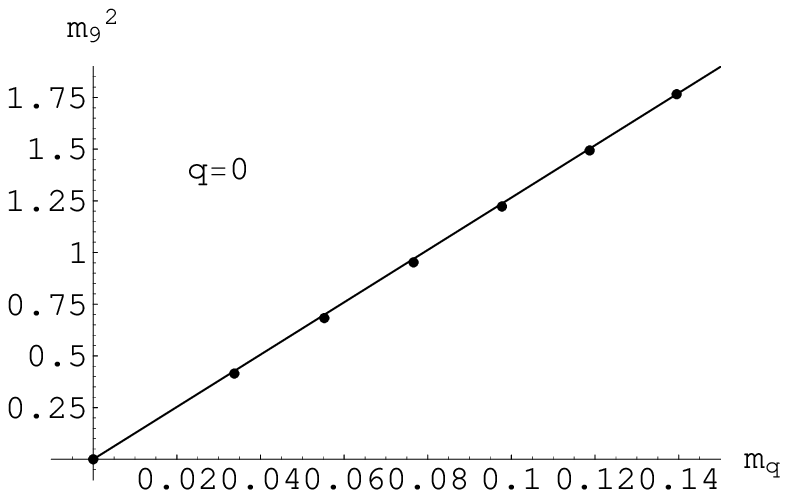} \label{mass3-fig}}
\subfigure[] {\includegraphics[angle=0,width=0.45\textwidth]{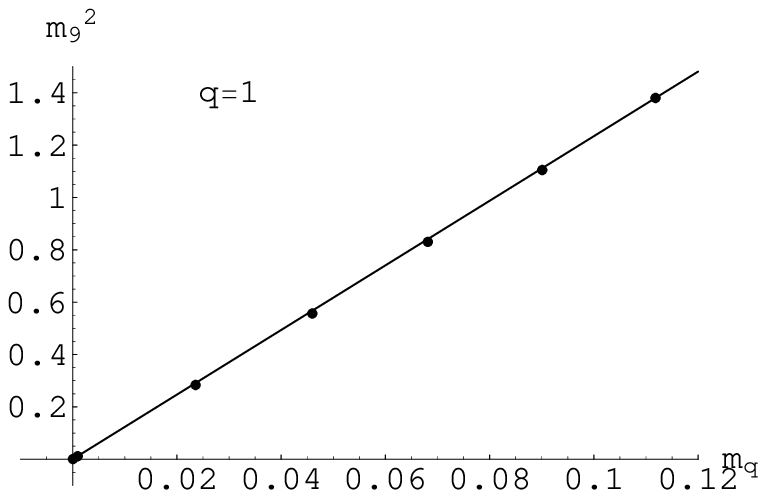} \label{mass4-fig}}
\caption{$m_9^2$ vs $m_q$ for $l=n=0$, $r_0=1.0$ and $R=1$ 
and (a) $q=0$, (b) $q=1$.
}
\end{center}
\end{figure}

In Figs.~\ref{mass-fig}, \ref{mass2-fig}, the numerical results of the mass
eigenvalues, $m_9$ and $m_8$, are plotted as functions of $m_q$.  
These values are all for the nodeless solutions, i.e. for the lowest
mass states.  It seems that they approach to the same values at large 
$m_q$ \cite{BGN}.  On the other hand, as seen from Figs.~\ref{mass3-fig}, 
\ref{mass4-fig}, we have verified that the relation $m_9^2\propto m_q$ 
holds very accurately in the small $m_q$ region for both $q=0$ and $q=1$.  
This fact establishes that $\phi_{l=0}^9$ is the Nambu-Goldstone boson 
as expected {and implies that some kind of relation like 
GellMann-Oakes-Renar.}  
                                                                                
Next, let us compare our numerical results with the mass formula
of scalars given in AdS$_4$ by Avis, Isham and Storey \cite{AIS}.  
In the Ref.~\cite{AIS}, the masses of scalars in AdS$_4$
are given
\beq
m_I^2= 
       K(I-3)I,\qquad I>2,  \label{ads4-mass}
\eeq
where $m_I$ is written for the case without the conformal coupling
of the scalar and the gravity, and 
$I$ is given by (i) $2<I<5/2$ or (ii) $I=3,4,5,\dots$.
The case (i) represents continuous spectrum while the case (ii) does
discrete one. The parameter $K$ is related to our $\lambda$ as,
\beq
  K={}^{(4)}R/12=-\lambda,  \label{K}
\eeq
where ${}^{(4)}R$ denotes four dimensional scalar curvature.

\begin{figure}[htbp]
\vspace{.3cm}
\begin{center}
\subfigure[$m_9^2$ vs $n,l$ at $m_q=10^{-5}$ with $r_0=1.0$, $R=1$ and $\lambda=-4$] 
{\includegraphics[angle=0,width=0.45\textwidth]
{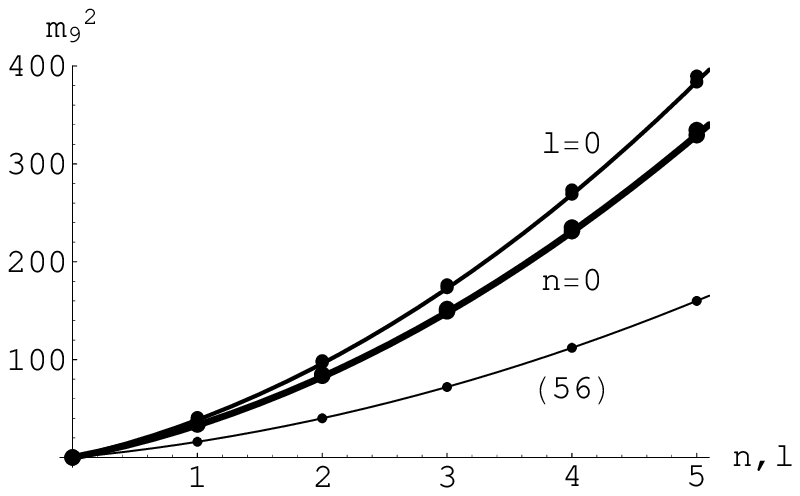} \label{mass5-fig}}
\subfigure[$m_8^2$ vs $n,l$ at $m_q=10^{-5}$ with $r_0=1.0$, $R=1$ and $\lambda=-4$] 
{\includegraphics[angle=0,width=0.45\textwidth]
{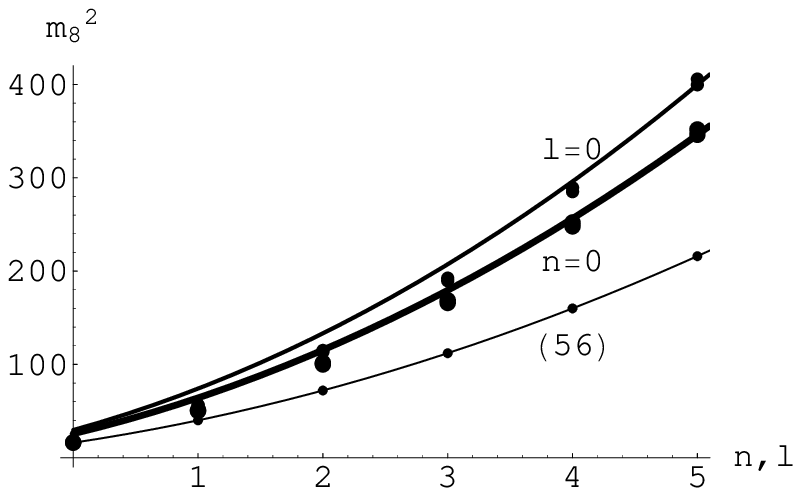} \label{mass6-fig}}
\caption{Small points and the thin curve corresponds to the formula 
(\ref{ads4-mass}).  Mid-large points and the mid-thick curve represent 
the obtained values and a fitted curve, respectively, for node ($n$) 
excitations with $l=0$ while large points and thick curve represent 
obtained values and a fitted curve, respectively, for $l$-excitations 
with $n=0$.  
} \label{Mass-fig1}
\end{center}
\end{figure}

Let us begin with eigenvalues at $m_q=10^{-5}$ shown in Fig.~\ref{Mass-fig1}.  
Since obtained eigenvalues have zero or finite nodes (denoted by $n$) 
or integer $l$, they correspond to the case (ii) of the mass formula 
(\ref{ads4-mass}). The continuous spectrum (i), which corresponds to 
tachyons, does not exist in our case.  It seems somewhat curious from the 
viewpoint of AdS/CFT correspondence that the corresponding spectrum to (i) 
does not exist.  And we consider that the ground 
state for $m_9^2$ corresponds to $I=3$ since it is massless while the 
ground state for $m_8^2$ corresponds to $I=4$ since it is massive.  
Obtained values are very good agreement with the formula (\ref{ads4-mass}) 
for the ground states.  Their values are $m_9^2=1.31\times 10^{-4}$ for
$q=0$ and $1.12\times 10^{-4}$ for $q=1$ while $m_8^2=16.0$ for $q=0$ 
and $16.5$ for $q=1$.  

\begin{figure}[htbp]
\vspace{.3cm}
\begin{center}
\subfigure[$m_9^2$ vs $n$ at $m_q=5.0$ with $r_0=1.0$, $R=1$ and $\lambda=-4$] 
{\includegraphics[angle=0,width=0.45\textwidth]
{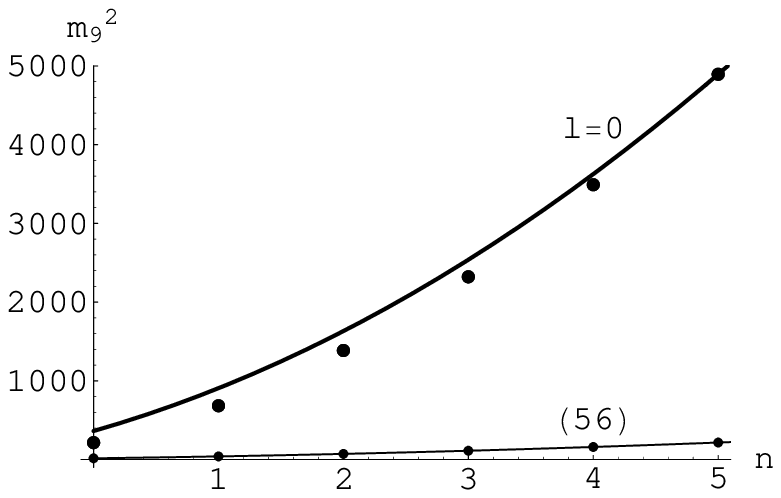} \label{mass7-fig}}
\subfigure[$m_8^2$ vs $n$ at $m_q=5.0$ with $r_0=1.0$, $R=1$ and $\lambda=-4$] 
{\includegraphics[angle=0,width=0.45\textwidth]
{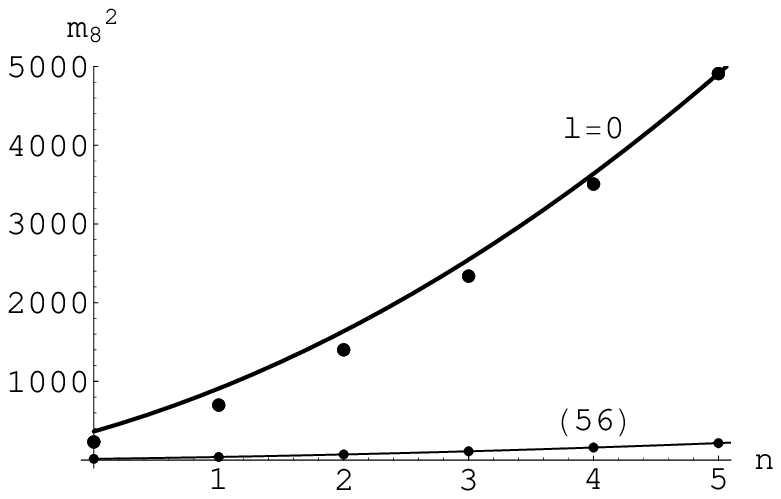} \label{mass8-fig}}
\caption{Points and the curve represent the obtained values and 
a fitted curve, respectively, for node ($n$) excitations with $l=0$.  
The lower lying points and the thin curve correspond to the mass 
formula (\ref{ads4-mass}).
} \label{Mass-fig2}
\end{center}
\end{figure}

\vspace{.3cm}
However, the numerical results deviate from the mass formula for the exited
states with respect to $n$ and $l$. We can see that
the larger $n$ ($l$) with $l=0$ ($n=0$), the 
deviations becomes larger from the formula (\ref{ads4-mass}).  { 
These deviations might be reduced to
the gauge interactions between quark and anti-quark which
make a bound state. As shown in the previous sub-section, 
the formula (\ref{ads4-mass}) is reproduced when we set $w=0$.
On the other hand, the mass deviates from this formula when we
take into account of non trivial $w$.
Since the profile function
$w$ includes the quark mass and the vacuum expectation value of 
$\bar{\Psi}\Psi$ and other information of the interaction between the
quarks and the gauge fields. Then, the deviation observed is naturally
understood as the reflection from the gauge interactions to the meson mass.
We can see this effect through the change of the mass scale 
in the mass formula as shown below. 
}

Fortunately obtained values are almost degenerate for $q=0$ and $q=1$ 
as seen from Fig.~\ref{Mass-fig1}.  So, we attempt to fit the data 
at the present stage by changing $K$ by hand such that the obtained 
values for the highest excited states with $q=0$ are reproduced.  
The purpose to fit in such a way is as follows; since obtained 
values for ground states are in good agreement with the mass formula 
(\ref{ads4-mass}) and deviate for higher excited states, we want to 
know conversely deviations for lower lying states when the above 
mentioned fitting are performed.  The results are as follows for 
$n$-excitations with $l=0$, 
\bea
&K_n^{(0)}=2.40K,\qquad\hbox{for $m_9^2$,} \cr
&K_n^{(0)}=1.85K,\qquad\hbox{for $m_8^2$,} \nonumber
\eea
and for $l$-excitations with $n=0$,
\bea
&K_l^{(0)}=2.06K,\qquad\hbox{for $m_9^2$,} \cr
&K_l^{(0)}=1.60K,\qquad\hbox{for $m_8^2$.} \nonumber
\eea
The coincidence is very (rather) good for $m_9^2$ ($m_8^2$) so that 
we may conclude that $I$-dependence of the formula (\ref{ads4-mass}) is 
trustable.  But it should be noted that $n$-excitations and $l$-excitations 
is split because $K_n^{(0)}$s and $K_l^{0}$s take different values.   
Although the mass formula (\ref{mass-formula}) is degenerate for 
$n$-excitations and $l$-excitations, it is obtained for $w=0$, 
which is not the real solution.  The real solution is $w\ne 0$ 
for $m_q=0$ and the formula (\ref{mass-formula}) does not contradict 
with the splitting mentioned above.

Next we study the case of large $m_q$.  
In the case of $m_q=5.0$, obtained values are rather large compared 
with the ones given by the mass formula (\ref{ads4-mass}) as shown 
in Fig.~\ref{Mass-fig2}.  This is because $K=-\lambda$ 
is independent of $m_q$ whereas we interpret it 
{should be changed as
$K_{(n,l)}\to \tilde{K}_{(n,l)}(\lambda, m_q, g_{\rm YM}^2,\dots)$.  }
But the definite and detailed form of 
$\tilde{K}$ 
is unknown at the 
present stage so that we again choose the fitting by changing $K$ 
by hand as $m_q=10^{-5}$ case.  The results are as follows for 
$n$-excitations with $l=0$, 
\bea
&K_n^{(5)}=22.7K,\qquad\hbox{for $m_9^2$,} \cr
&K_n^{(5)}=22.7K,\qquad\hbox{for $m_8^2$.} \nonumber
\eea
The coincidences are relatively good so that we may again conclude that 
$I$-dependence of the formula (\ref{ads4-mass}) is trustable.  
To derive more realistic mass formula is a future problem.  

\subsection{Baryon}

It has been shown  
that baryons correspond to D5-branes wrapped around the 
compact manifold $M_5$ \cite{Gross,Witten}. 
Here we assume it to be $S^5$. The brane action of 
such a D5 probe is 
\beq
S_{\rm D5}= -\tau_5 \int d^6\xi e^{-\Phi} \sqrt{\cal G} \ ,
\label{D5-action-2}
\eeq
where $(\xi_i)=(X^0,X^5 \sim X^9)$, $\tau_5$ represents 
the tension of D5 brane, and 
${\cal G}=-{\rm det}({\cal G}_{i,j})$ for the induced metric 
${\cal G}_{ij}= \partial_{\xi^i} X^M\partial_{\xi^j} X^N G_{MN}$. 
The mass of the wrapped D5-brane is then 
\bea
M_{\rm D5}(r,\lambda)=\tau_5 e^{-\Phi} \sqrt{\cal G} =
\tau_5 \pi^3 R^4 r A(r) e^{\Phi/2} \; .
\eea
Before seeing the $\lambda$ dependence of this quantity, we consider the
case of $\lambda=0$, 
\beq
M_{\rm D5}(r,0)=\tau_5 \pi^3 R^4 r \sqrt{1+ {\tilde{q} \over r^4}} \; .
\eeq
where $\tilde{q}$ is given by (\ref{gc}).
This has a global minimum at $r=r_{\rm min}=\tilde{q}^{1/4}$. Its
value is given as 
$M_{\rm D5}(r_{\rm min})=\tau_5 \pi^3 R^4 (4\tilde{q})^{1/4}$, 
and this is regarded as the baryon mass here.
In the 4d Minkowski limit, namely at
$\lambda=0$, the baryon mass is 
induced by the $\tilde{q}$, i.e. by the gauge-field condensate. 

\begin{figure}[htbp]
\vspace{.3cm}
\begin{center}
\subfigure[ 
The curves show $M_{\rm D5}|_{\lambda=0}$, 
$M_{\rm D5}|_{\lambda=-4}$, and $M_{\rm D5}|_{\lambda=-5.76}$, 
respectively, from bottom to up for $\tilde{q}=1$. ]
{\includegraphics[angle=0,width=0.45\textwidth]
{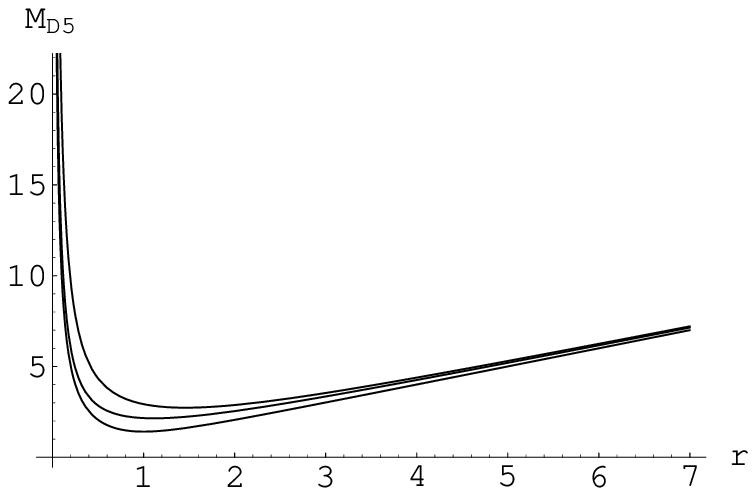} \label{D5massro}}
\subfigure[For $\lambda =-4$, 
$M_{\rm D5}|_{\tilde{q}=0}$, 
$M_{\rm D5}|_{\tilde{q}=10}$, and $M_{\rm D5}|_{\tilde{q}=20}$, 
are shown from bottom to up.] 
{\includegraphics[angle=0,width=0.45\textwidth]
{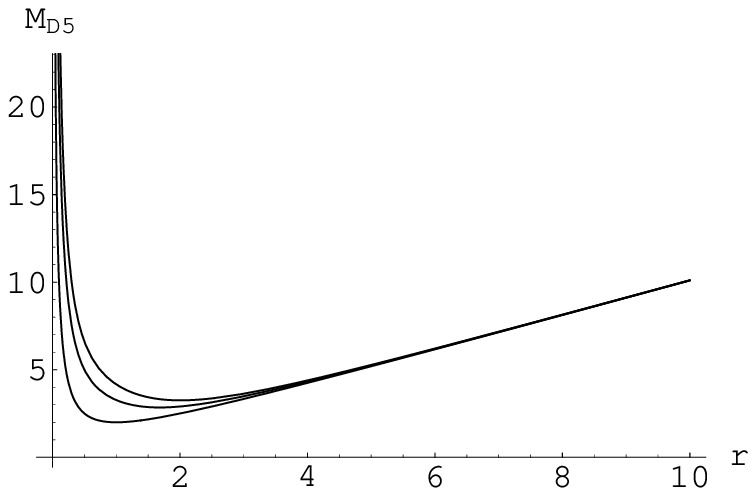} \label{D5massq}}
\caption{ The D5-brane mass $M_{\rm D5}(r)$ as a function of $r$
for $R=1$ and $\tau_5=1/\pi^3R^4$.
} \label{Mass-fig8}
\end{center}
\end{figure}

In the presence of $\lambda$, namely in AdS${}_4$, the minimum of
$M_{\rm D5}$ appears even if $q=0$. For $q=0$, we obtain 
$M_{\rm D5}(r=r_0,0)/(\tau_5 \pi^3 R^4)=2r_0$ as the minimum, which
is realized at $r=r_0$.

Fig.~\ref{Mass-fig8}(a) and (b) 
show the $r$ dependence 
of $M_{\rm D5}(r)$ for three values of $\lambda$ and $q$ respectively.
For any finite values of $\lambda$, even if $q=0$, there exist 
a minimum at an appropriate point of r, and the minimum value of $M_{\rm D5}$
increases with $\lambda$ and $q$. This point is consistent with
the above analysis of mesons, whose mass increases with these parameters.


\section{Summary}
In this paper, the non-perturbative properties of the gauge theories 
in the AdS${}_4$ are studied in the dual supergravity by including 
the light flavor quarks, which are introduced by a D7 brane embedding.

In our model, in the limit of Minkowski space limit ($r_0\to 0$), we have
$ g_{\rm YM}^2=1+{\tilde{q}\over r^4}$,
which is equivalent to the supersymmetric solution given previously \cite{GY}.
In this limit, quark is confined due to the strong YM coupling constant
in the infrared limit, $r\to 0$, for $\tilde{q}>0$. On the other hand,
in the AdS${}_4$ space-time (finite $r_0$), we find
$ g_{\rm YM}^2\to 1+{q\over 3}$ in the infrared limit, $r\to 0$.
While this coupling constant seems to be small to get the quark confinement,
we find the confinement by calculating the Wilson loop which leads to the 
area law and the divergent effective single quark mass. 
These results are obtained even if $q=0$, and the important factor to
get the confinement is reduced to the 
warp factor $A(r)$ which diverges at $r=0$ for finite $r_0$. 
In this sense,
contrary to the case of the Minkowski space-time, 
the gauge interaction does not play any important role
for the quark confinement in the AdS${}_4$ case. The important point
is the geometry itself.

\vspace{.3cm}
As for the chiral symmetry, we could expect its spontaneous breaking
due to the repulsion between the D7 and the D3 branes.
Actually we could find a numerical solution for the profile function $w(\rho)$,
with $m_q=0$ and $c>0$, and it has a finite D7 energy which is obtained
after $\rho$ integration.
On the other hand, the trivial solution $w(\rho)=0$, the one of $m_q=c=0$,
leads to an infinite D7 energy due to the infinite repulsion at $r=0$ due
to the D3 branes. We thus find the spontaneous breaking of the chiral symmetry.

In the present case, 
for all the solutions of $w(\rho)$, we need a correction like
$m_q\lambda\log(\rho)$ to the expectation value of $\la\bar{\Psi}\Psi\ra$.
This is expected when conformal and supersymmetry is broken even 
in the UV limit as in the present model. Since this correction is proportional
to $m_q$, the loop corrections of the massive hypermultiplet added to CFT would
be responsible to this result.

\vspace{.3cm}
The scalar meson spectra are calculated and we could 
assure the Goldstone boson due to the chiral symmetry breaking. 
The coincidence of the obtained masses to 
the mass formula given in Ref.~\cite{AIS} are very good
for the lowest states. 
On the other hand,
for the exited states, our results deviate from the mass formula.
However, when we modify
$K$ by hand in the mass formula 
(\ref{ads4-mass}), we find better coincidences. This is interpreted
as the reflection of the gauge interaction between quark and anti-quark
when they form bound states.

And baryon mass is studied by regarding it as the energy of the 
D5 brane, wrapped on $S^5$, embedded at a stable point with respect
to the coordinate $r$. For any finite values of 
$\lambda$, even if $q=0$, we find a minimum or a stable point at 
an appropriate point of $r$. This is consistent with the meson case.

\vspace{.3cm}
\section*{Acknowledgments}

This work has been supported in part by the Grants-in-Aid for
Scientific Research (13135223)
of the Ministry of Education, Science, Sports, and Culture of Japan.


\newpage
\end{document}